\documentstyle[12pt,aasms4]{article}
\begin{document}

\title{Asymmetry measures for QSOs and companions\footnote{Based on 
observations obtained with the Canada France Hawaii Telescope, which 
is operated by CNRS of France, NRC of Canada, and the University of Hawaii}}

\author{J.B. Hutchings, C. Proulx }

\affil{Herzberg Institute of Astrophysics, 5071 West Saanich Rd.,
Victoria, B.C. V9E 2E7, Canada; john.hutchings@nrc.ca}

\begin{abstract}

An asymmetry index is derived from ellipse-fitting to galaxy images, that 
gives weight to faint outer features and is not strongly redshift-dependent. 
These measures are made on a sample of 13 2MASS QSOs and their neighbour
galaxies, and a control sample of field galaxies from the same wide-field
imaging data. The QSO host galaxy asymmetries correlate well with 
visual tidal interaction indices previously published. The companion 
galaxies have somewhat higher asymmetry than the control galaxy sample, 
and their asymmetry is inversely correlated 
with distance from the QSO. The distribution of QSO-companion asymmetry 
indices is different from that for matched control field galaxies at 
the $\sim95\%$ significance level. We present the data and discuss this 
evidence for tidal and other disturbances in the vicinity of QSOs.

\end{abstract}


\section{Introduction and data}

   QSO host galaxies have been associated with tidal disturbances and
signatures of merger events, and it appears that such events are often
the triggers for nuclear activity (see e.g. reviews by Veilleux 2006 and 
Hutchings 2006). Investigating red QSOs discovered by 2MASS, Hutchings
et al (2003, 2006) find that these have more pronounced signs of interaction
than blue QSOs, and suggest that the red population is in an earlier stage
of interaction where associated star-formation produces dust that reddens 
the active nucleus. 

   In these investigations, we used visual estimates of interaction level
from the images. While these are clearly seen, it is not simple to
quantify the interaction signatures. In this paper we introduce and use
an asymmetry measure made on the images, which allows us to make more
quantitative statements on the galaxies.

   In addition to the host galaxies themselves, it has been speculated 
that neighbouring galaxies may be disturbed, either by outflow or
radiation effects from the QSO, or because QSOs are found in environments
where there are other merging or interacting galaxies. In this paper we
report asymmetry measures on galaxy neighbours of 13 of the 2MASS
QSOs in the redshift range 0.3 to 0.6 (from Hutchings et al 2006). 
As a control set we have made the same measures on field galaxies of similar magnitude, `surrounding' 
random stars in the same imaging data as the QSOs. In addition, the
asymmetry index was measured for a number of stars in the fields, to
provide the zero point for asymmetry for each observation.

   The data are described by Hutchings et al (2006) and are from the CFHT
Megaprime camera with the r-band filter. The pixel scale is 0.187 arcsec.
Most fields had several exposures, and these were measured individually.
Table 1 identifies the QSOs (by their RAs in Hutchings et al 2006),
with some measured quantities. Note that there were 43 companion galaxies
in total, but 122 individual measurements were made as the fields were all
observed more than once. The control galaxies were selected from the same
observations so that the image quality was matched.

\section{Asymmetry index}

  The most characteristic sign of tidal events or merging in galaxies
is asymmetry of the morphology. These events produce faint arms that
are seen outside the central galaxy, which in time fall back into
the final galaxy configuration, but are visible for periods that probably 
are longer than a QSO episode. Disk warps and patchy dust may also occur.
In the early stages of a merger, the bright central parts of both galaxies may 
be seen. The common property of all these is that the galaxy does not
have symmetrical structure, so that an objective way of measuring
a lack of symmetry is a valuable tool in looking for such events.
Many such indicators are seen in the faint outer parts of galaxies, so 
that such a measure should weight these appropriately compared with the 
bright inner part.

   The IRAF task `ellipse' fits ellipses to image data contours, and produces 
a table of quantities for each ellipse. In particular, for each contour 
level ellipse, it reports a semi-major axis (SMA) and ellipticity, and
deviations  from this ellipse in the data. To derive an overall measure 
of the asymmetry of an object, we want to average these deviations over 
a range of signal level contours. This average should be weighted in
a way that gives the appropriate significance to faint outer features
which reveal global asymmetries, 
compared with the much larger signal from the innermost pixels, which 
may be more affected by image quality and detector sampling. 

   Other authors have used an asymmetry measure which is derived from
180$^o$ rotation of a galaxy image, subtracted from the original (e.g. 
Conselice et al 2003, Lotz et al 2004, and references therein). As noted
above, we wish to add weight to the faint outer features that are signs of 
tidal events, rather than flux-weighted asymmetry of a whole galaxy. We also
want a measure that can be compared over different redshifts (i.e. size and 
flux differences) since we are dealing with a range of redshifts
in this work. 

   After experimentation with a number of recipes, the following
was adopted for this investigation: for each contour level, 
asymmetry = contour signal level x (sum of absolute values of third
harmonic deviations from ellipse) x SMA$^{1.5}$.  The SMA is
measured in units of image pixels. These asymmetry values
may be averaged over a range of contours to yield a total asymmetry
value for the galaxy. To enable comparison between galaxies with 
different brightness and size on the sky, this mean value is divided by the 
total signal in the galaxy. This is to ensure that the same galaxy with an 
intrinsic asymmetry, will have the same total asymmetry index 
at different distances (redshifts). 

   The innermost contours in the ellipse task may be oversampled and 
also subject to PSF differences and detector saturation. To avoid such
effects in the summed asymmetry index, we do not include values for radii
less than 5 pixels (about 0.9 arcsec). Far away from the galaxy, the flux 
falls and the index may be affected by noise, bright pixels, or small
unrelated objects in the line of sight. Thus we do not use contours that
lie beyond where the contour flux falls below 10\% of the average contour 
level signal to that radius value.  

 It is of course necessary to edit out stars or hot
pixels that do not belong to the object. Accurate sky-subtraction is 
important for this measure, so that the asymmetry 
index eventually approaches zero far from the galaxy centre. To compare 
between different galaxies, the ellipse contour sampling should be 
done using the same radius steps in pixels, and the numbers  
normalised to the total signal value as noted above. Radial scaling is not
done as the weighting by radius makes the effect of different
redshift small. Thus, to compare galaxies with this index, we do not need
to know their redshifts.

   Figure 1 shows the effect of moving a galaxy to higher redshifts.
This shrinks the image linearly with redshift and reduces the signal level
by the inverse square of the redshift, for the low redshift range of interest. 
The plot shows the contour asymmetry values with scaled radius, after
normalising to the total signal, at double and triple the redshift.
The asymmetry and its variation with radius are all very similar for the
three plots, so that we may compare asymmetry values meaningfully over 
the range of redshifts of interest in the sample. This type of plot of
individual contour asymmetry with ellipse semi-major-axis, also shows 
where the principal asymmetries lie within the galaxy. 

   Another potential measure of asymmetry in faint outer parts of a
galaxy is the wandering of the ellipse
centroids as the SMA increases. We looked at the standard deviation of
the centroid coordinates for an image as the simplest way to code this,
but it had no correlation at all with the asymmetry measure above, or
with the visual index from Hutchings et al (2003,2006). Perhaps a more
complex measure of systematic centroid wandering, weighted by contour signal,
would be better, but this approach was not explored any further, since the
above formula works well, and correlates well with the visual estimates
of asymmetry in our earlier paper on the same objects.

\section{Measurements}

    Measurements of asymmetry as described above were made on the program
QSOs and galaxies near to them. Galaxies comparable with or fainter than 
the QSO were
measured, out to a radial distance of $\sim$320 pixels (1 arcmin). A number
of stars were also measured as controls for the low limits for asymmetry 
from each image, and for the consistency of sky-subtraction.  In view of 
the range of image quality and depth we do not claim to have a complete 
sample to faint limits, and this forms part of the discussion below.

   The main control sample was to measure a similar ensemble of galaxies
`surrounding' randomly selected stars in the same images, of brightness similar
to the QSOs. Their distances from the central star were recorded and
this sample extended to about 450 pixels. Noting the difference in
the QSO companions and control galaxies, it is likely that there are
more galaxies around QSOs than around a random place in the sky. 

   To compare with the observations, we constructed a simple model
in which the mean asymmetry index varies from 0 to 200 linearly with distance 
from a QSO in the range 0 to 400Kpc, with a spread of about 50. 
This was sampled with points roughly as the volume of space defined by 
the separation, and with several random realisations of
projections on the plane of the sky. We discuss the comparison below. 

   We also made a model table with asymmetry indices and distances having
the same spreads as the QSO companion measures. From this table we generated
many randomised distributions of asymmetry with distance, to compare with 
the measured distribution. This too is discussed below.

   Table 2 shows mean asymmetry indices for the different classes of
object in this work, including the model outlined above.  

   In comparing the QSO asymmetries with those of the galaxies, 
we note that the QSO total signals include the bright nuclei, 
so these should be removed in comparing the flux-normalised asymmetry 
indices, as in Table 2. The nuclear to host ratios for
the QSOs in the sample are taken from Hutchings et al (2006), and
reduce the total fluxes by an average factor 5.
These host galaxy fluxes are still several times brighter than their
average companion galaxies, and 50\% brighter than the brightest.
The comparison between the galaxy companions 
and control galaxy samples are more robust, since they do not contain
bright AGN, and are the main interest of the discussion below. 

  The asymmetry indices for all the QSOs were derived independently by
the two authors. Hutchings et al (2006) published
`interaction indices' for the QSOs, based on visual inspection of the
images for signs of tidal disturbances. Indeed, the visible appearance
of interacting galaxies was used in developing the asymmetry index.
Figure 2 shows the comparison of the aymmmetry values from both the
authors, plotted against the interaction index. In deriving the
asymmetry values, each of us derived the sky subtraction and
decided on editing of bad pixels etc, independently. The agreement 
between the two sets of values indicates the level of spread or uncertainty
that we should attach to them. The good general correlation with 
the interaction index also indicates that we have a useable objective
measure of tidal disturbance. The values plotted in Figure 2 are the
mean asymmetry indices without correction for the total and nuclear 
fluxes, so are higher than those for the galaxies. As the nuclear fluxes
are variable and not well defined, such corrections add extra scatter
to the plot, although the correlation with visual interaction
index is still clear. 

The dots in Fig 2 are the author (CP) measures used for all other objects 
in this paper. They show a higher spread than the other (JBH) QSO measures 
but the QSOs are the most difficult objects to measure because of the
high nuclear flux in many of them. The overall spread with interaction index
indicates that the measure does miss some of the visual signs of tidal
disturbance (or possibly that the visual signs are overinterpreted).

One of the QSOs is at much higher redshift (2.37), so that we do
not expect to detect companions or measure their morphology at the same
level as the others, since they would be several magnitudes fainter.
We have excluded it from the discussion below, but
it is useful as a further control on the main results. The QSO has
very low asymmetry index and so do most of its `companions'.

\section{Discussion}

   Figure 3 shows the asymmetry measure plotted against image quality,
redshift, and distance from an arbitrary `central' star for the control 
sample. None of these shows a strong correlation, envelope, or dependence.
High values of the asymmetry measure are seen more in better image conditions
(top panel) so we looked at these separately to see if any
indications are different. In practice, excluding the 23 galaxy measures
with FWHM above 1.1" does not alter the distribution of the
asymmetry measures with separation from the QSOs. There no  
indication that asymmetry is lower in the higher redshift object 
companions (panel 2), but we note that since these are line of sight 
companions, more may not have the QSO redshift in the higher redshift cases.
Nevertheless, eliminating the higher redshift objects does not change 
the distribution of companion galaxy properties with projected distance 
from the QSOs.

   Figure 4 shows the distribution of nearby galaxy asymmetries with 
projected distance from the QSOs, in the top panel. This distribution is 
characterised by an upper envelope that drops with increasing distance from
the QSO. The distances are converted to Kpc using H$_o$=75. The figure
shows the full set of measures for all the observations. The smaller
dots represent 1/3 of the points that are matched in asymmetry
distribution with the control galaxy sample, as possible elimination of
line of sight companions that are not associated with the QSO. We
discuss below the differences in these distributions. Table 2 shows
the mean properties of these as `decontaminated'. There are no 
companions brighter than the host galaxy of the QSOs (see Hutchings 
et al 2006). As noted, the distribution is not signifcantly
altered by eliminating QSO fields with redshift about 0.4, or data
with FWHM less than 1.2 arcsec, as noted in the discussion of Fig 3.

   In Figure 4 we also compare the measures with the simple model described 
in the section above. The true distribution is shown in the centre panel, and
the combination of 5 random realisations of projection angles is shown in
the bottom panel. The model assumes that we have sampled all galaxies
evenly by volume of space, out to about 200Kpc and then more sparsely 
beyond that. This corresponds with what we did measure - we have not 
measured all of the galaxies in the outer parts of the distribution 
because of image defects or overlaps. We have consistently measured 
galaxies down to the same flux limits, in both QSO and control fields,
at all distances from the central QSO or star. It is also clear 
from Figs 3 and 4 that
we find no galaxies with large interaction indices at large distances from
the QSO, while this is not so for the control galaxies.

   Overall, the data are very consistent with the model, and thus
suggestive that there may be some disturbing effect of the QSO on 
nearby galaxies, or that QSOs are associated with tidal events in 
fairly dense groups of galaxies. 

   We may compare the asymmetries of the control galaxies to see if they
differ from the QSO companions. Looking at the distribution of
asymmetry values, the K-S test indicates that they are from different
populations at the 91\% probability level. If we restrict the control sample
to having matched pixel distances or magnitudes to the companions sample, 
this value goes to 92\% and 93\%, respectively. Finally, if we compare the
companion galaxies with 1/3 removed as line of sight superposed field 
galaxies with the same distribution as the control
galaxy asymmetries (Fig 4 top panel and Table 2), the distributions are
different at the 95\% level. If the non-companion fraction is higher than
1/3, the probability of them being different populations is still higher. 
These numbers thus also support the idea that the QSO companions are 
more disturbed than field galaxies with similar magnitudes. 

   We also compared the mean asymmetry values for QSO companions with those
for the control galaxies, in distance bins of 50 pixels. The QSO companion
asymmetries are larger than the controls, at all radii, but differ most
significantly in the 50-100 radius pixel bin. The averages are 125 for the
QSO companions and 68 for the control sample. If we decontaminate the
companions with 1/3 having the properties of the control group, the
mean value for the companions rises to 149, and the difference becomes
3.1 times its formal uncertainty. These and other numbers are given in Table 2. 

   Finally, from many random combinations of the distributions of
distance and asymmetry, we find less than 5\% cases with the falling
distribution seen in the top panel of fig 4. Thus the envelope of
asymmetry with distance as observed is very unlikely to be a chance 
occurrence.  

   The density of companion galaxies measured around QSOs is quite high. We
measured galaxies that are down to 10\% of the QSO host galaxy average.
This is an absolute R magnitude in the range -21.6 to -19.1, so
we are measuring companions down to the luminosity of the LMC. This galaxy
density amounts to 0.3 within 50Kpc, 1.6 within 100Kpc, 2.8 within 200Kpc,
and 3.6 within 400Kpc, assuming that all are associated with the QSO. 
The counts are less complete beyond about 200Kpc, but
these numbers are more dense than the local group. The
control galaxy counts were not intended to be complete, although areas of
sky rich in galaxies were chosen for convenience. Even so, the density
of galaxies in the control sample is several times lower than around the
QSOs. In such dense groups, tidal interactions are likely to be common,
again supporting the finding that the asymmetry is higher close to the QSOs.

   Further work on this would include colour information on the
galaxies, to see if there is a difference in stellar population age, or
dust in the QSO companions. The apparent finding that the companions are
most disturbed nearest the QSO suggests that either the QSOs reside
in the central (and brightest) galaxy of a group, or that the QSO is causing 
the disturbances we measure in their companions. Given that these are
red QSOs, which are obscured and likely to be new, the former of these
seems the more likely scenario. 

   We thank the referee for helpful comments. 


\clearpage

\begin{deluxetable}{ccccccc}
\tablenum{1}
\tablecaption{Summary of sample objects}
\tablehead{\colhead{Name} &\colhead{\#obs} &\colhead{z} &\colhead{r-mag} 
&\colhead{\#galaxies} &\colhead{Galaxy mags} &\colhead{FWHM"} }
\startdata
1332 &2 &0.346 &17.1 &3 &21.3-22.1 &0.86-1.01\cr
1432 &2 &0.349 &16.0 &3 &20.9-21.4 &0.82-0.84\cr
1435 &2 &0.305 &16.3 &3 &20.9-21.4 &0.90-0.95\cr
1442 &2 &0.307 &17.1 &4 &20.3-21.8 &0.86-0.92\cr
1450 &4 &0.358 &17.0 &3 &21.1-21.6 &0.64-0.69\cr
1501 &2 &0.337 &18.3 &3 &21.1-22.1 &0.90\cr
1549 &2 &2.37 &17.0 &3 &20.2-21.9 &0.65-0.69\cr
1550 &2 &0.373 &16.6 &3 &19.8-20.3 &0.64\cr
1618 &2 &0.446 &18.2 &3 &20.3-20.6 &0.84-0.97\cr
1644 &10 &0.329 &18.3 &3 &20.1-21.1 &1.2-1.4\cr
1700 &4 &0.596 &19.7 &5 &19.7-21.3 &1.0-1.1\cr
1700 &4 &0.509 &16.7 &4 &18.3-20.8 &0.73-0.79\cr
1715 &2 &0.524 &19.8 &3 &21.8-22.2 &0.80-0.86\cr
\enddata
\end{deluxetable}

\begin{deluxetable}{lcccc}
\tablenum{2}
\tablecaption{Mean properties of program objects}
\tablehead{\colhead{Objects} &\colhead{Number} &\colhead{Mean asymmetry} 
&\colhead{Av r-mag} &\colhead{Mean sep in Kpc}}
\startdata
13 QSOs  &42 &17 &17.3 &0\cr
QSO hosts &-- &85 &18.8 &0\cr
Comp galaxies &122 &81 &20.5 &141\cr
Comp decontaminated &84 &89 &20.9 &138\cr
Comp decont 50-100 px radius &10 &149 &21.3 &56\cr
PSF stars &105 &9 &17.7 &--\cr
Control galaxies &136 &74 &21.0 &(130)\cr
Control galx 50-100 px radius &9 &68 &21.0 &(58)\cr
Model &20 &78 &-- &150
\enddata
\end{deluxetable}


\centerline{\bf Captions to Figures}

\figcaption{Asymmetry index with radius for a well-resolved asymmetric barred 
spiral galaxy (solid line). The innermost values, which oversample the 
bright nuclear regions, are not plotted. The galaxy has
been artifically removed to 2 and 3 times the distance by reducing the 
signal and pixel binning (dotted and dashed lines respectively). The
asymmetry index is relatively free of redshift systematics.  }

\figcaption{QSO asymmetry measures and visual interaction indices from Hutchings 
et al (2006). The dots are the measures made in this work and the circles 
are measures made earlier and independently by one of us (JH)
on the whole Hutchings et al (2006) sample. The line 
connects the mean values for each index value.  The values are not 
scaled to the mean flux of the sample galaxies or by the estimates of 
the nuclear flux, in order to show only the
agreement between the independent sets of measures, and the correlation 
with the visual interaction index.}

\figcaption{Asymmetry measures which may reveal biases. Top: QSO companion (dots)
galaxies and control (circles) galaxies with image quality. Middle:
Companion galaxies (dots) and total-flux corrected QSOs (stars) with 
redshift of the QSO. Bottom: Control
galaxies with distance from arbitrarily chosen central star.} 

\figcaption{Distribution of asymmetry measures with projected distance from QSO.
Top: measured program objects. Small dots are those that
would disappear if 1/3 of them are not associated with the QSO and
have the asymmetry distribution of the control galaxies in Figure 3.
Middle: model with asymmetry correlated with distance from QSO, as sampled. 
Bottom: 5 realisations of 
random projections of the model from the middle panel. The similar
distributions in the top and bottom panels show the data are
consistent with the model. }

\clearpage

\plotone{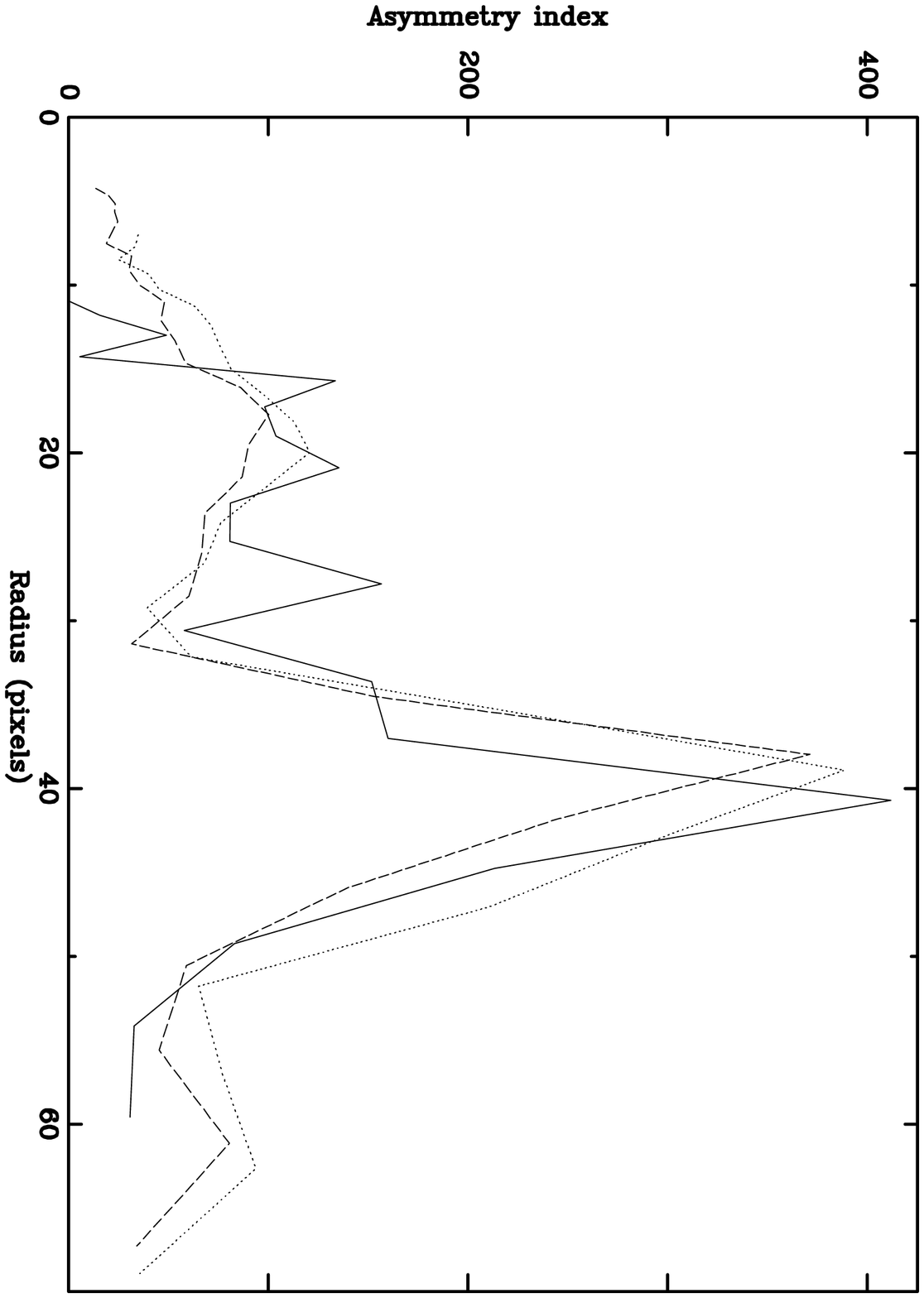}
\newpage
\plotone{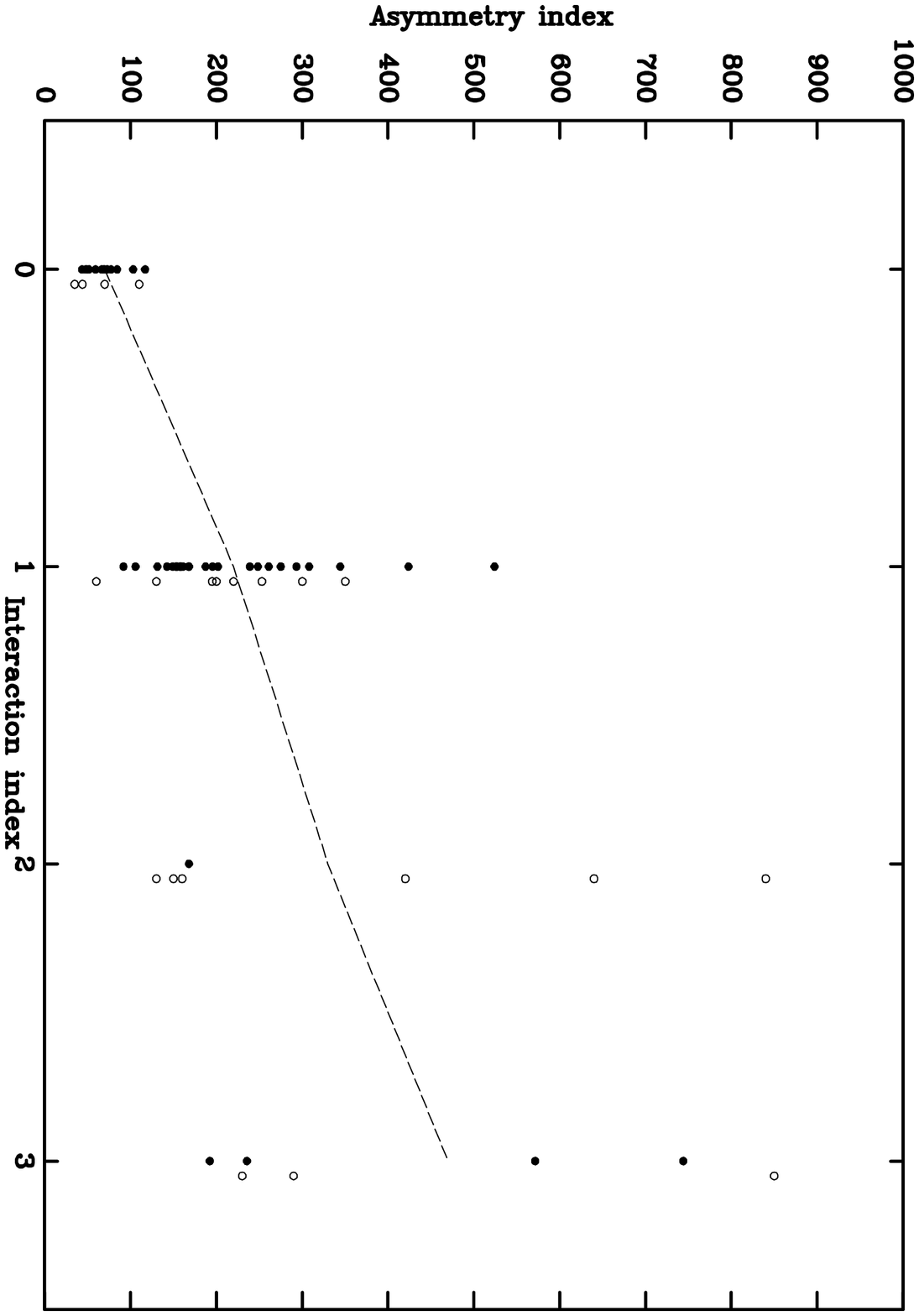}
\newpage
\plotone{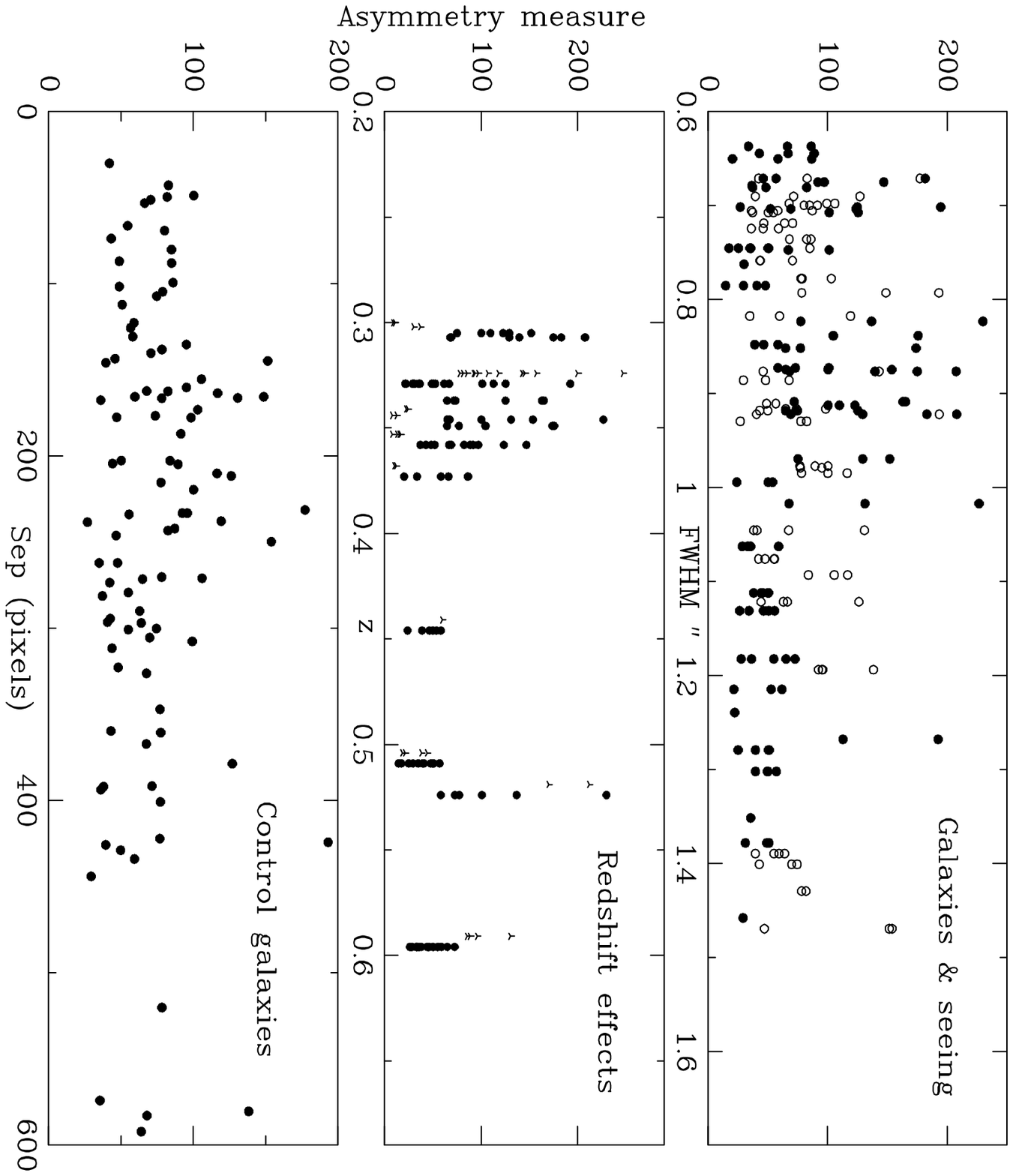}
\newpage
\plotone{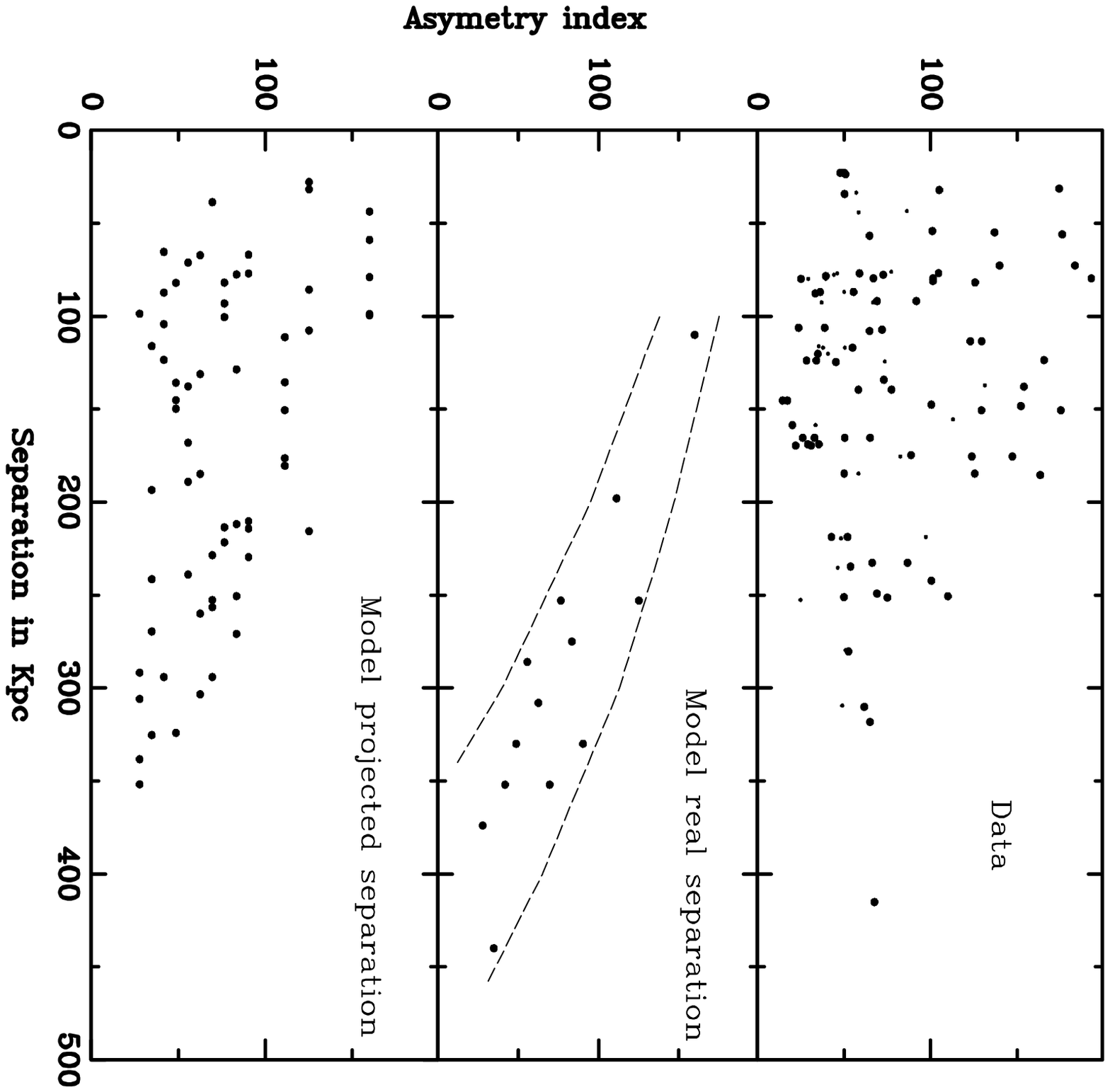}

\end{document}